\documentclass[journal=jacsat,manuscript=article]{achemso}

\usepackage[version=3]{mhchem} 
\usepackage[utf8]{inputenc}
\usepackage{amsmath}
\usepackage{siunitx}
\usepackage{graphicx}

\usepackage[hidelinks]{hyperref} 
\usepackage[nameinlink]{cleveref}

\newcommand{\siliconnitride}{SiN$_\mathrm{x}$ }
\newcommand{\alumina}{Al$_2$O$_3$ }
\newcommand{\micron}{\micro\meter}

\author{Yufan Li}
\affiliation[QN]
{Department of Quantum Nanoscience, Kavli Institute of Nanoscience, Delft University of Technology, Delft 2628CJ, The Netherlands}
\alsoaffiliation[3me]
{Department of Precision and Microsystems Engineering, Faculty of Mechanical Engineering, Delft University of Technology, Delft 2628CD, The Netherlands}

\author{Gesa Welker}
\affiliation[QN]
{Department of Quantum Nanoscience, Kavli Institute of Nanoscience, Delft University of Technology, Delft 2628CJ, The Netherlands}

\author{Richard Norte}
\affiliation[QN]
{Department of Quantum Nanoscience, Kavli Institute of Nanoscience, Delft University of Technology, Delft 2628CJ, The Netherlands}
\alsoaffiliation[3me]
{Department of Precision and Microsystems Engineering, Faculty of Mechanical Engineering, Delft University of Technology, Delft 2628CD, The Netherlands}

\author{Toeno van der Sar}
\affiliation[QN]
{Department of Quantum Nanoscience, Kavli Institute of Nanoscience, Delft University of Technology, Delft 2628CJ, The Netherlands}
\email{t.vandersar@tudelft.nl}

\title[An \textsf{achemso} demo]
  {A robust, fiber-coupled scanning probe magnetometer using electron spins at the tip of a diamond nanobeam}

\begin{document}







\begin{abstract}
  Fiber-coupled sensors are well suited for sensing and microscopy in hard-to-reach environments such as biological or cryogenic systems. We demonstrate fiber-based magnetic imaging based on nitrogen-vacancy (NV) sensor spins at the tip of a fiber-coupled diamond nanobeam. We incorporated angled ion implantation into the nanobeam fabrication process to realize a small ensemble of NV spins at the nanobeam tip. By gluing the nanobeam to a tapered fiber, we created a robust and transportable probe with optimized optical coupling efficiency. We demonstrate the imaging capability of the fiber-coupled nanobeam by measuring the magnetic field generated by a current-carrying wire. With its robust coupling and efficient readout at the fiber-coupled interface, our probe could allow new studies of (quantum) materials and biological samples.  \\

\end{abstract}

\section{Introduction}

Visualizing magnetic phenomena with high spatial resolution has proven to be a powerful tool in both fundamental physics and applied sciences. Magnetometry with nitrogen-vacancy (NV) centers in diamond stands out for its nanoscale imaging capability \cite{Balasubramanian2008NanoscaleConditions,Maze2008NanoscaleDiamond,Maletinsky2012ACentres}, operability from cryogenic to above room temperature, and its large range of application areas\cite{Casola2018ProbingDiamond,Schirhagl2014Nitrogen-vacancyBiology,LeSage2013OpticalCells,Steinert2013MagneticResolution,deGroot2021MicromagneticRock-Magnetism}. Advances in NV magnetometry in the past decade have led to breakthroughs especially in revealing the nanoscale physics of condensed matter systems  \cite{Casola2018ProbingDiamond,Gross2017Real-spaceMagnetometer,Thiel2019ProbingMicroscopy,Sun2021MagneticImaging,Simon2021DirectionalWaves}. However, the free-space optics generally used for optical interrogation of the NV spins are challenging to realize in cryogenic, intra-cellular, or other hard-to-reach environments. As such, realizing robust all-fiber-based NV probes with efficient optical readout could enable new measurements in low-temperature (quantum) or biological systems. 

Achieving high spatial resolution and sensitivity in scanning-probe imaging requires the sensor to be in close proximity to the target sample, to have a small size, and to provide efficient addressing and readout. Fiber-coupled NV probes based on diamond microcrystals placed onto the end of cut fibers have enabled micron-scale magnetic imaging \cite{Fedotov2014Fiber-opticImaging,Chatzidrosos2021FiberizedMagnetometers,Dix2024ACenters}. However, reaching higher spatial resolution requires confining the NV sensor spins to a nanoscopic volume located at the tip of a scanning-probe assembly. 

Here, we realize a fiber-coupled diamond nanobeam sensor with an ensemble ($\sim$1000) of NV spins deterministically implanted into the nanobeam tip. By attaching the beam to a tapered optical fiber using ultraviolet-curing optical glue \cite{Parker2024ARegister}, we realize a robust and transportable nanobeam-fiber assembly with optimized optical coupling efficiency. We demonstrate through-fiber microscopy by monitoring the photoluminescence at probe-sample contact, enabled by the robust nature of the assembly. The through-fiber detected photoluminescence enables navigation over the sample surface, control over the fiber-sample distance, and scanning-probe magnetometry without free-space optics. We demonstrate the magnetic imaging capabilities of the probe by measuring the magnetic field generated by an omega-shaped current-carrying wire on a chip.

\section{Results and Discussion}

\subsection{Fabrication and assembly of the fiber-coupled diamond nanobeam}

\begin{figure}[hbt!]
    \centering
    \includegraphics[width=\textwidth]{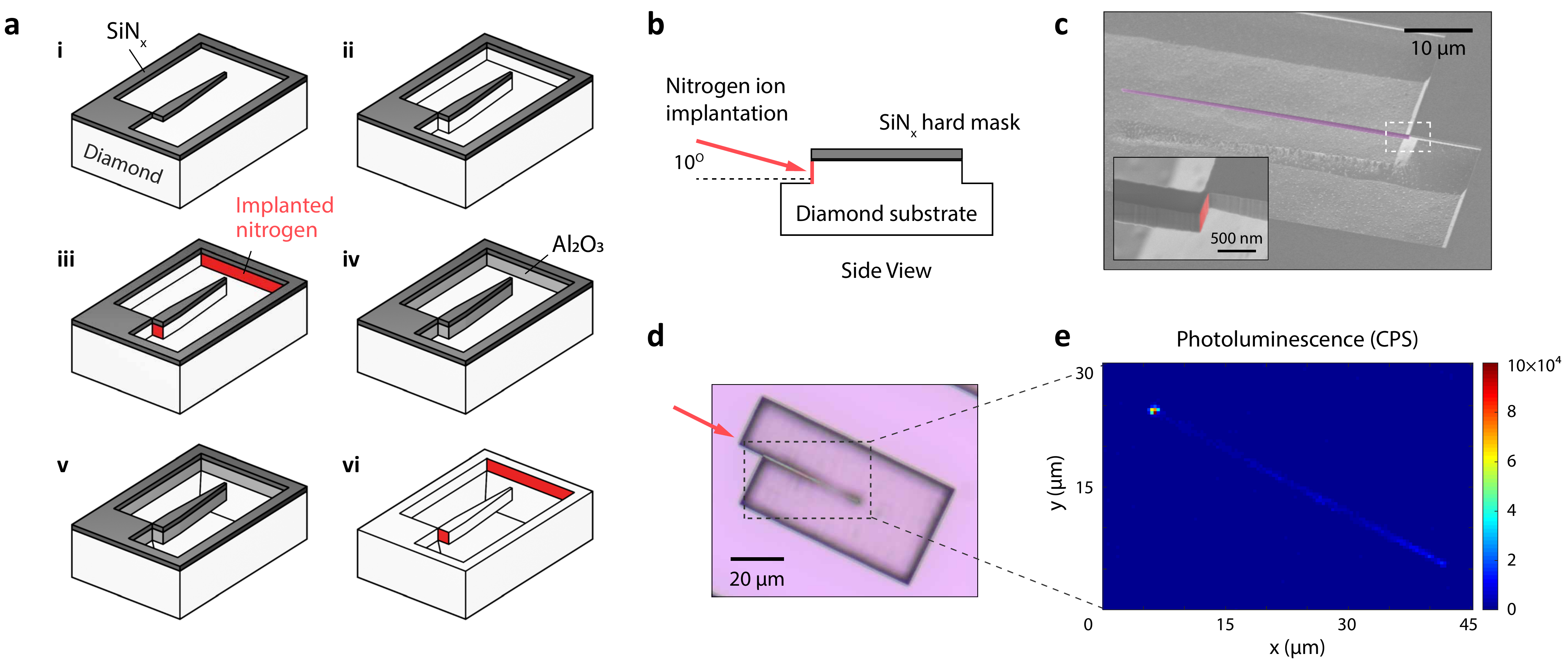}
    \caption{\textbf{Fabrication of diamond nanobeams with deterministically implanted NV centers at the nanobeam tip. (a)} Fabrication flow: (i) Deposition and patterning of a \siliconnitride hard mask (dark grey) on a single-crystal diamond substrate (white); (ii) Anisotropic diamond etch to transfer the mask to the diamond and create the nanobeam sidewalls; (iii) Nitrogen ion implantation at 80 degrees with respect to the surface normal, parallel to the beam direction and into the end facet, as illustrated in (b). The sidewalls facing the implantation are marked in red; (iv) Sidewall protection with \alumina (light grey); (v) Quasi-isotropic diamond etch to undercut the nanobeams. (vi) Removal of all mask materials with hydrofluoric acid. \textbf{(b)} Side-view schematic of the angled implantation. \textbf{(c)} SEM image of the undercut diamond nanobeam (false-colored in purple) after fabrication. The nanobeams are aligned along the $\langle110\rangle$ orientation of the diamond lattice. The inset shows a zoomed-in image of the nitrogen-implanted end facet, false-colored in red. \textbf{(d)} Optical microscope image of a diamond nanobeam, with the red arrow indicating the implantation orientation. \textbf{(e)} Photoluminescence map in the black dashed rectangle area in (d), measured by scanning a focused green laser over the diamond substrate using a home-built scanning confocal microscope. The bright spot indicates NV centers concentrated at the end of the nanobeam as expected.}
    \label{fig:fab}
\end{figure}

We fabricate the diamond nanobeams using a combination of anisotropic and isotropic reactive ion etching (Methods) developed in ref. \cite{Pasini2023NonlinearWaveguide}. To realize diamond nanobeams with NV centers at their tips, here we incorporate angled ion implantation into the fabrication flow (\cref{fig:fab}a). We start with an electronic-grade, single-crystal diamond (Element-Six, nitrogen concentration $<\SI{5}{ppb}$) and have $^{14}$N ions implanted (INNOViON Corp., \SI{50}{keV}, \SI{1e13}{\per\cm\squared}) after the anisotropic diamond etch that defines the beam sidewalls (\cref{fig:fab}a, iii). To obtain NV centers at the nanobeam tips, we implant the ions at a grazing angle of \SI{10}{\degree} (\cref{fig:fab}b) into the end facets while the top surface of the beam is shielded by the \siliconnitride hard mask. Given the $500\times\SI{500}{\nm\squared}$ surface area of the nanobeam end facet, we expect $\sim2.5\times10^4$ implanted nitrogen ions at the nanobeam tip. SRIM simulations\cite{Ziegler2013SRIM-2013} predict an average implantation depth of \SI{60}{nm} underneath the nanobeam end facet.

To realize free-hanging NV-nanobeam sensors, we undercut the nanobeams via isotropic reactive ion etching (\cref{fig:fab}c, d) \cite{Mouradian2017RectangularDiamond,Pasini2023NonlinearWaveguide,Li2023ANanobeam}, remove the mask materials via acid cleaning, and then vacuum anneal at \SI{800}{\celsius} to form NV centers (see Methods). We confirm the formation of an NV ensemble at the nanobeam tip by imaging the nanobeam photoluminescence with a home-built scanning confocal microscope. (\cref{fig:fab}e). A distinct bright spot indicates the desired presence of the NV sensor spins at the end of the beam.

\begin{figure}[hbt!]
    \centering
    \includegraphics[width=\textwidth]{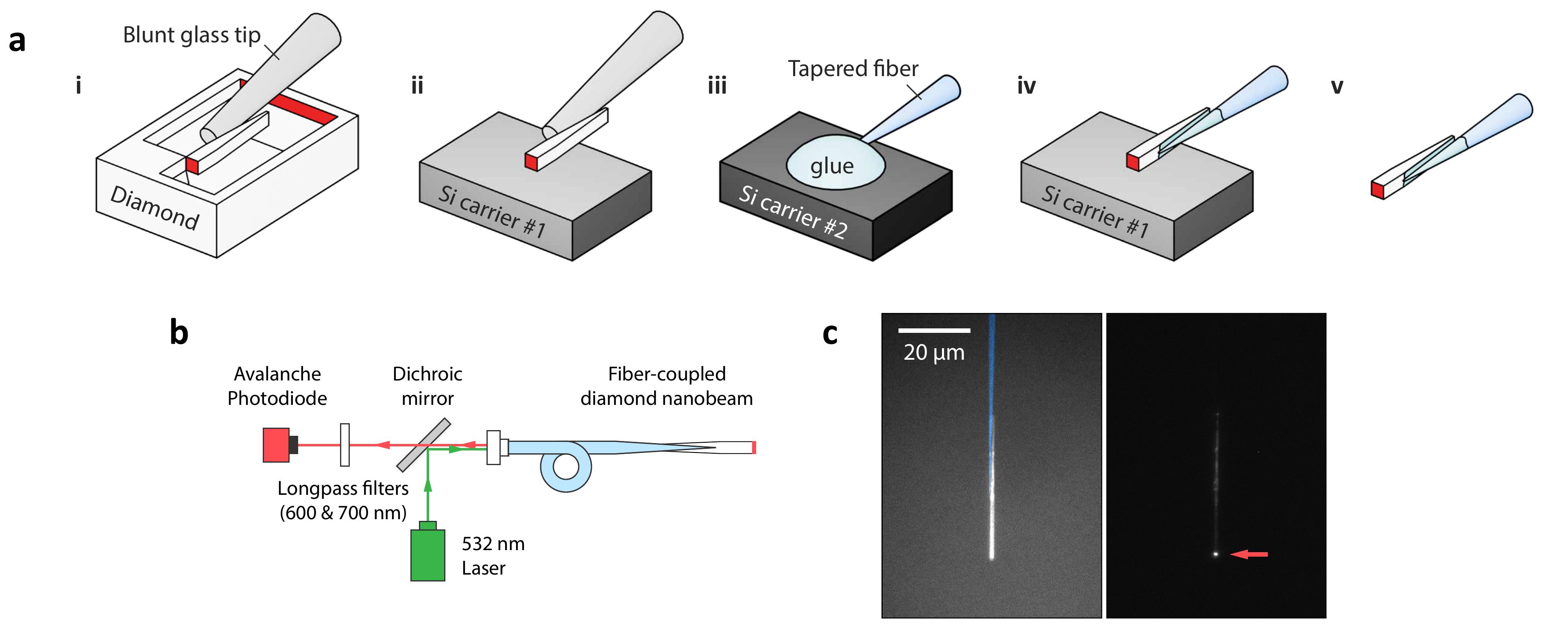}
    \caption{\textbf{Assembling the fiber-coupled nanobeam sensor with optimized photon collection efficiency. (a)} Assembly workflow using optical glue: (i) A blunt glass tip is used to break the nanobeams off the diamond substrate; (ii) The blunt tip places the nanobeam at the edge of a silicon carrier; (iii) Optical glue is applied to the tapered fiber by dipping the fiber tip into a glue droplet on another carrier chip, and then retracting the fiber; (iv) The glue-covered tapered fiber is brought into contact with the nanobeam on the edge, and the glue is cured with UV illumination after optimizing the coupling between the fiber and the nanobeam; (v) The glued fiber-nanobeam is retracted from the carrier edge. \textbf{(b)} Simplified schematics of the optical setup. Two long-pass filters at \SI{600}{\nm} and \SI{700}{\nm} are placed in front of the photodiode to reduce background photoluminescence. \textbf{(c)} Microscope image of the glued fiber-nanobeam assembly under ambient illumination (left, the fiber is false colored in blue) and with through-fiber \SI{532}{\nm} laser excitation (right, with the red arrow indicating the end of the beam). Both images are taken with a \SI{650}{\nm} long-pass filter in front of the camera to block the excitation laser. }
    \label{fig:assemble}
\end{figure}

Attaching the nanobeam robustly to the fiber is crucial for realizing a sensor that is transportable to different setups and that does not break upon probe-sample contact. Here we attach the nanobeam to the fiber using a thin layer of optical glue (Norland Optical Adhesive 86H) \cite{Parker2024ARegister}. To do so, we mount the diamond with the free-hanging nanobeams (\cref{fig:fab}c) onto a slip-stick positioner (Mechonics MX-35) and bring a nanobeam into contact with a blunt glass tip (tip radius $\sim\SI{5}{\micron}$) attached to a holder (fig. S1). We drive the positioner such that the tip pushes the nanobeam sideways until it breaks off and sticks to the tip (\cref{fig:assemble}a, i). We then remove the diamond substrate, replace it with a silicon carrier chip, and temporarily place the nanobeam on the chip edge (\cref{fig:assemble}a, ii). 

To apply the glue, we replace the blunt tip with the tapered fiber and the nanobeam carrier chip with a chip carrying a small droplet of glue. We dip the fiber into the droplet and then pull it out at a speed of about \SI{1}{\micron/\second} (\cref{fig:assemble}a, iii). This speed ensures that a thin layer of glue forms on the fiber without excessive droplets, ready for attaching the nanobeam.  

We glue the fiber to the nanobeam in a way that aims to maximize the optical coupling efficiency: We re-mount the nanobeam-carrier chip and bring the glue-covered fiber into contact with the nanobeam lying on the chip edge (\cref{fig:assemble}a, iv). At this stage, the large contact area between nanobeam and carrier causes the beam to remain stuck to the carrier chip. This enables adjusting the relative position between fiber and nanobeam while monitoring the NV photoluminescence that we excite and detect through the fiber (\cref{fig:assemble}b). When the detected photoluminescence is maximal, we fix the fiber position and illuminate the fiber-nanobeam assembly with a UV lamp (Thorlabs CS2010) to cure the glue. After curing, the bonding between the fiber and the nanobeam is sufficiently strong to overcome the adhesion to the carrier, such that we can detach the glued assembly (\cref{fig:assemble}a, v). White-light and through-fiber-excited-photoluminescence images show the resulting fiber-coupled nanobeam sensor with the NV centers embedded at the tip (\cref{fig:assemble}c).

\subsection{Characterization of through-fiber NV photoluminescence readout}

\begin{figure}[hbt!]
    \centering
    \includegraphics[width=\textwidth]{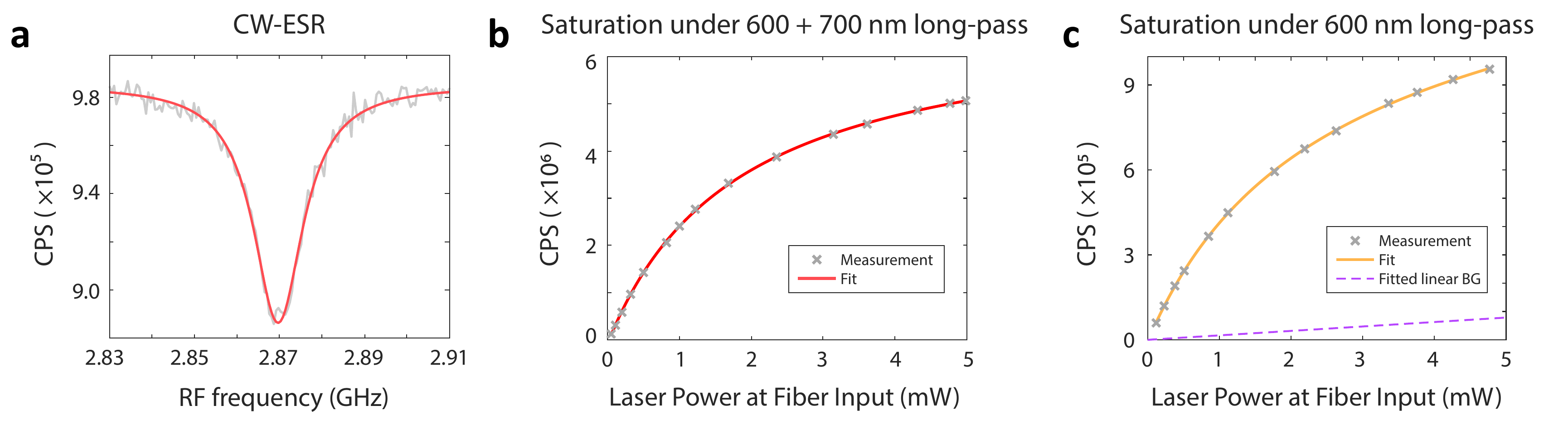}
    \caption{\textbf{Characterizing the through-fiber NV photoluminescence readout. (a)} Electron spin resonance (ESR) signal of the NV centers in the nanobeam measured through the fiber, measured with 30 µW of laser excitation (measured in free space in front of the fiber coupler) at zero external magnetic field, under \SI{600}{\nm} + \SI{700}{\nm} long-pass filtering. A Lorentzian fit (red curve) yields a contrast of 10.0(2)\% and a \SI{7.1(3)}{MHz} zero-field full width half maximum. \textbf{(b)} Optical saturation of measured photoluminescence, measured under \SI{600}{\nm} + \SI{700}{\nm} long-pass filtering. Additional neutral density filtering of ND1.5 is applied before the APD, in order for the PL to stay within the measurement range of our APD. Measured data (grey cross) are fitted with \cref{eq:sat} (red curve). \textbf{(c)} Optical saturation measured under \SI{600}{\nm} long-pass filtering, and the fit using \cref{eq:sat} (orange curve). Additional neutral density filtering of ND3 is applied. The non-zero linear background extracted from the fit is plotted as the purple dashed line.}
    \label{fig:charac}
\end{figure}

We characterize the electron spin resonance (ESR) spectrum of the fiber-coupled nanobeam sensor by exciting and detecting the NV photoluminescence through the fiber (\cref{fig:assemble}b). We drive the ESR by applying a microwave current through a stripline on a chip positioned $\sim\SI{50}{\micron}$ below the nanobeam. We find that adding a \SI{700}{\nm} long-pass filter in the detection path yields a 10.0(2)\% ESR contrast (\cref{fig:charac}a), which is 60\% larger than when filtering with only a \SI{600}{\nm} long-pass filter (see SI). This indicates a strong background photoluminescence in the $600\sim\SI{700}{\nm}$ range.

To determine the origin of this background, we measure the photoluminescence intensity $I$ as a function of laser excitation power $P$ under two filtering conditions: \SI{600}{\nm} + \SI{700}{\nm} long-pass filters for the first measurement (\cref{fig:charac}b), and a single \SI{600}{\nm} long-pass filter for the second measurement (\cref{fig:charac}c). We fit both datasets to the model
\begin{equation}
\label{eq:sat}
    I(P) = I_\mathrm{sat}\frac{P}{P+P_\mathrm{sat}} + kP + I_0.
\end{equation}
Here, the first term models the optical saturation of NV centers \cite{Dreau2011AvoidingSensitivity}, characterized by a saturation power $P_\mathrm{sat}$ and an asymptotic intensity $I_\mathrm{sat}$. The second term accounts for background photoluminescence, produced by e.g.\ the fiber or the glue, which we assume to be linear with laser power with proportionality constant $k$ \cite{Li2023ANanobeam}. The final term $I_0$ represents the dark count rate of our detector. 

For the \SI{600}{\nm} + \SI{700}{\nm} filtered measurement, the fit yields $k = \SI{-0.01(7)e6}{\per\second\per\milli\watt}$, or $k = 0$ if we bound $k$ to $k>0$ (\cref{fig:charac}b). We conclude that the detected fiber or glue autoluminescence above \SI{700}{\nm} is insignificant compared to the NV luminescence, indicating potential for reaching the single-spin regime. For the \SI{600}{\nm} filtered measurement, the fit yields $k = 0.016\pm\SI{0.015e6}{\per\second\per\milli\watt}$ (\cref{fig:charac}c). As such, the ratio between the NV signal and the background is still $I_\mathrm{sat}/kP_\mathrm{sat} \approx 37 \gg 1$ for the low powers $P \ll  P_\mathrm{sat}$ used in our measurements. Therefore, this small linear background PL cannot explain the observed difference in ESR contrast between the two filtering regimes (\cref{fig:charac}a and SI). Combined with further analysis on the measured ESR contrasts (SI), we conclude that the background photoluminescence originates predominantly from neutral NV centers in the diamond nanobeam. Increasing the NV$^-$ to NV$^0$ ratio, for instance by diamond surface treatments \cite{Zheng2022CoherenceBath}, is therefore of primary importance for future improvements of our fiber coupled sensors.

\subsection{Demonstration and analysis of scanning magnetometry}

\begin{figure}[hbt!]
    \centering
    \includegraphics[width=\textwidth]{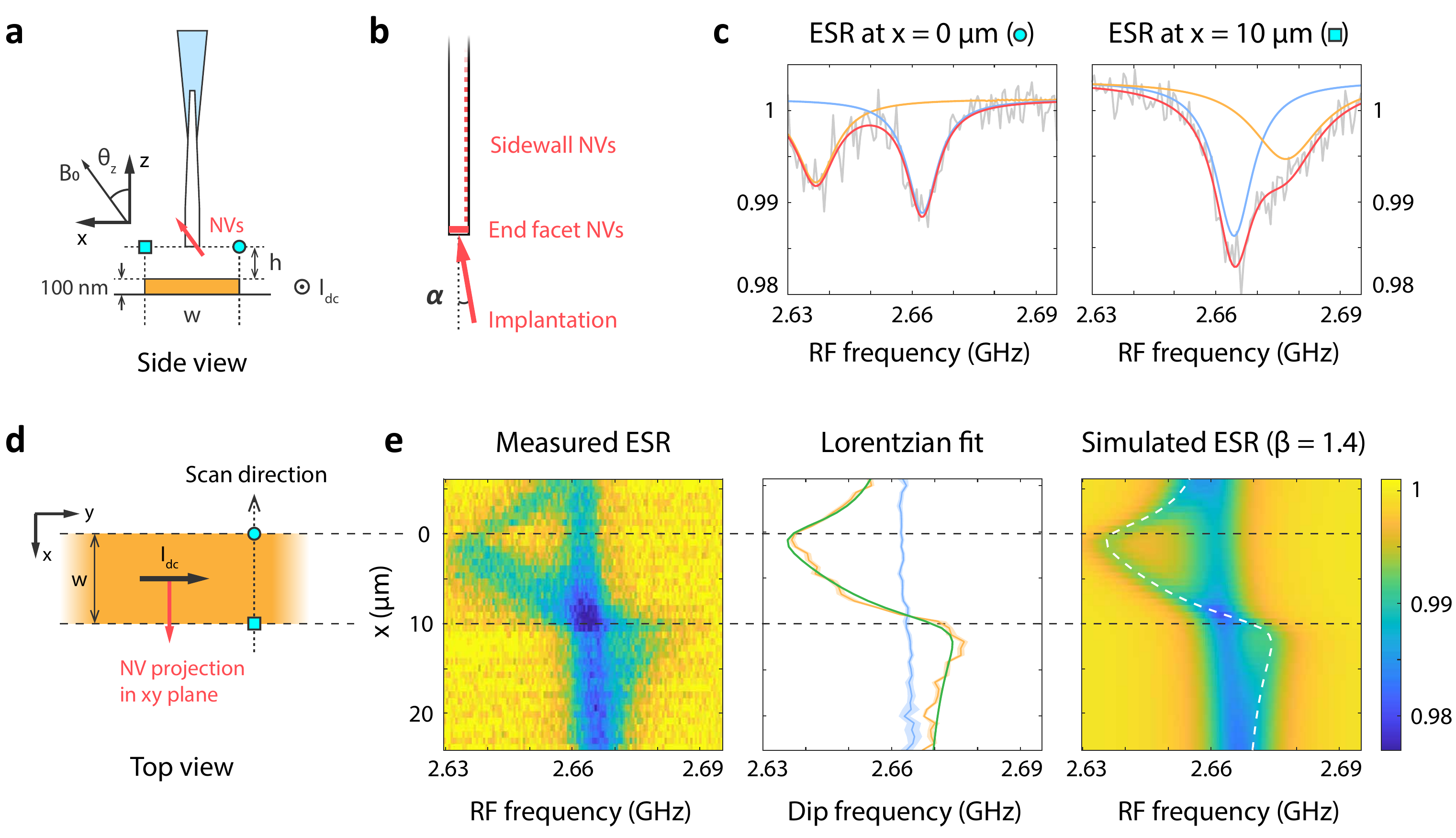}
    \caption{\textbf{Characterizing the NV distribution in the nanobeam by imaging the magnetic field of a current-carrying wire. (a)} Side view of the measurement scheme. The nanobeam is scanned above a \SI{100}{nm}-thick, \SI{10}{\micron} wide gold strip on a silicon substrate. $I_\mathrm{dc} = \SI{20}{mA}$ is applied through the gold strip along with the microwave signal used to drive ESR. An external magnetic field $B_0=\SI{7.31(6)}{mT}$ is applied along one of the four crystallographic NV orientations, at \SI{35}{\degree} with respect to the beam axis ($\langle110\rangle$) and perpendicular to $y$ axis. We use the lower ESR frequency of this NV orientation in subsequent magnetometry measurements. \textbf{(b)} Illustration of how a small misalignment $\alpha$ during nitrogen implantation can lead to NVs distributed along the beam sidewall. \textbf{(c)} ESR spectrum measured at $x = \SI{0}{\micron}$ (left, cyan circle in (a)) and $x = \SI{10}{\micron}$ (right, cyan square in (a)). The data (light grey) is fitted with a double Lorentzian (red) to extract the contribution from the end-facet (orange) and sidewall NVs (blue). \textbf{(d)} Top view of the measurement scheme. \textbf{(e)} \textit{Left}: Normalized ESR spectra measured across the strip (dotted arrow in (d)). \textit{Middle}: ESR frequencies extracted via a double Lorentzian fit as in (c). We fit the extracted end-facet ESR frequencies (orange curve) using a model of the wire magnetic field (green). Free parameters are the offset and scaling in $x$, the lift height $h$ and NV orientation $\theta_z$ with respect to $z$ axis. The fit yields $h = \SI{1.8(2)}{\micron}$ and $\theta_z = \SI{43(3)}{\degree}$. Right: Calculated ESR line shape across the strip based on \cref{eq:esrspec}. The ESR frequencies at the end facet (white dashed curve, same as the green curve in the middle panel) are superimposed on the plot for reference.}
    \label{fig:misalign}
\end{figure}

We assess the magnetometry performance of our fiber-coupled nanobeam sensor by measuring the magnetic field generated by a DC electric current in a microstrip on a chip (\cref{fig:misalign}a, Methods). We mount the chip vertically with respect to the nanobeam and apply an external magnetic field $B_0$ along one of the four possible crystallographic orientations of the NV centers to isolate their ESR transition. To measure the field generated by the DC current, we first position the beam at about \SI{1}{\micron} above an edge of the strip. While we expected the ESR resonance to shift with respect to its value at $B_0$ because of the microstrip field, we surprisingly observed that the ESR dip splits (\cref{fig:misalign}c). By measuring the NV ESR spectrum in a line scan across the strip, we observe the evolution of this split peak in the changing strip field (\cref{fig:misalign}e). A strongly shifting dip and a weakly shifting dip are clearly visible.   

The observed splitting of the ESR response indicates the presence of two sub-ensembles of NV centers in our nanobeam that experience different magnetic fields. Assuming the strongly shifting dip corresponds to the NVs at the nanobeam tip (as these NVs should experience the largest magnetic field from the wire), we can fit the magnetic field extracted from this dip to the calculated wire field at a distance $h$ above the sample. We find an accurate match with the data for $h = \SI{1.8(2)}{\micron}$. From the same fit, we can extract the NV angle $\theta_z$ (\cref{fig:misalign}a), as this angle leads to a spatial asymmetry of the signal. We find $\theta_z = \SI{43(3)}{\degree}$. The difference from the \SI{35}{\degree} angle expected from the crystal orientation indicates a small tilt of the fiber-nanobeam with respect to the normal of the sample surface. We conclude that, despite the presence of a weakly shifting ESR resonance, we can accurately extract the surface magnetic field using the NV ensemble at the tip.

We attribute the weakly shifting resonance to NVs that were spuriously implanted into an unprotected beam sidewall because of a small misalignment during implantation (\cref{fig:misalign}b). SRIM simulations of \SI{50}{keV} nitrogen implantation at a $\alpha = \SI{1}{\degree}$ misalignment angle with respect to the diamond surface  yield an average implantation depth of $\sim\SI{10}{nm}$, sufficient to form a low-density, optically addressable NV layer along the \SI{40}{\micron}-long beam sidewall. To check the validity of this assumption, we model the measured ESR spectra $I(f,x)$ by integrating the ESR response of NV centers that are partially located at the end facet and partially distributed homogeneously along the beam sidewall:
\begin{equation}
\label{eq:esrspec}
    I(f,x) = \frac{1}{1+\beta}\left(I_0(f,B(x,h)) + \frac{\beta}{L}\int_h^{h+L}I_0(f,B(x,z))dz\right).
\end{equation}
Here, $I_0(f,B)$ is the normalized PL intensity of a single NV center, $B(x,z)$ is the magnetic field in the nanobeam, $h = \SI{1.8}{\micron}$ denotes the tip-sample distance determined previously, and $L = \SI{40}{\micron}$ is the length of the nanobeam. We assume all NV centers have equal contrast and linewidth. We weigh the relative contribution of the sidewall NVs by a factor $\beta$, which parametrizes their different number and coupling efficiency to the optical mode in the fiber. This model accurately reproduces the measured split-dip ESR spectra (\cref{fig:misalign}e right) for $\beta=1.4$. 

\subsection{2D microscopy and magnetometry using a fiber-coupled diamond nanobeam}

\begin{figure}[hbt!]
    \centering
    \includegraphics[width=\textwidth]{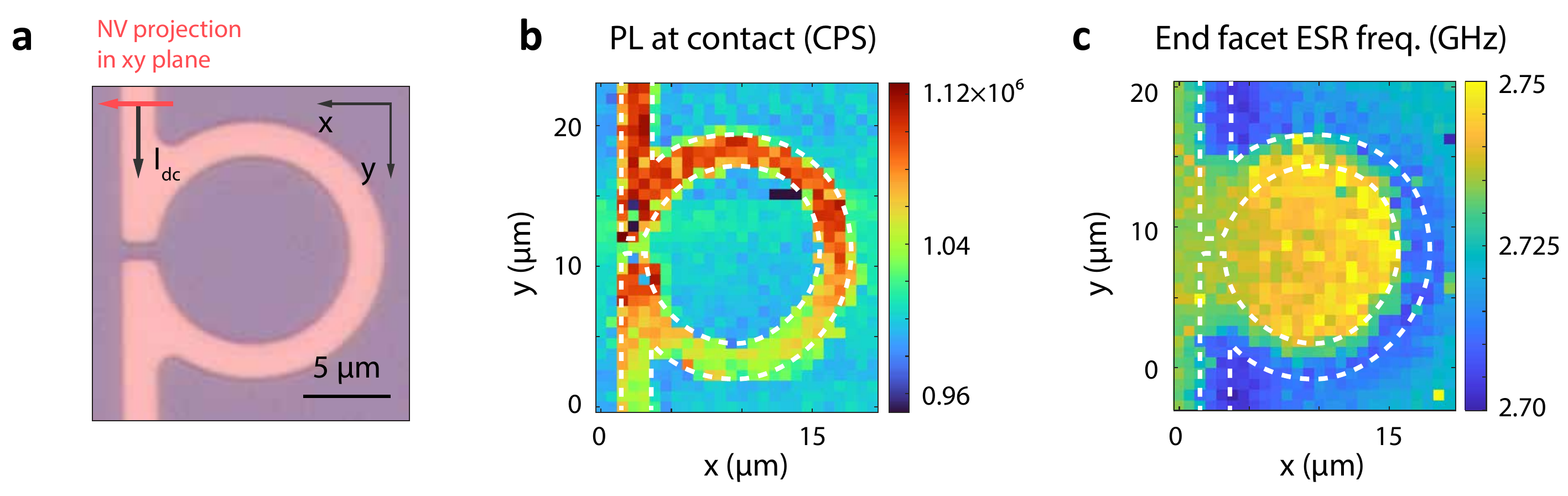}
    \caption{\textbf{Magnetic imaging with electron spins in a fiber-coupled diamond nanobeam. (a)} Optical microscope image of the omega-shaped gold strip. The width of the strip is \SI{2}{\micron}. \textbf{(b)} Through-fiber microscopy of the omega-shaped strip. At each pixel, the nanobeam is moved towards the sample. The increased photoluminescence upon contact is shown in the image. \textbf{(c)} 2D map of the end-facet ESR frequency, with $B_0 = \SI{5.0(2)}{mT}$, $I_\mathrm{dc} = \SI{10}{mA}$ and $h\sim\SI{1}{\micron}$. At each pixel, the end-facet and sidewall ESR frequencies are extracted via a double Lorentzian fit with the sidewall ESR frequency fixed at \SI{2.729}{GHz}. White dashed lines indicate the boundary of the gold strip inferred from (b). Both (b) and (c) are post-corrected for drift based on the known shape of the strip (see SI for the original drift-distorted image).}
    \label{fig:imaging}
\end{figure}

To demonstrate the 2D imaging capabilities of our fiber-coupled nanobeam, we characterize the magnetic field generated by a direct current that is sent through an omega-shaped, \SI{2}{\micron}-wide gold strip (\cref{fig:imaging}a). As our setup does not have atomic force microscope feedback, it requires an alternative method for maintaining a fixed tip-sample distance in the presence of thermal drifts and sample tilts. We do so by bringing the tip in contact with the sample at each pixel and then retracting the tip by a fixed amount. To detect tip-sample contact, we move the tip towards the sample until we observe a sharp increase in the detected photoluminescence (SI). We find that this increase is highly sensitive to the local sample reflectivity, enabling through-fiber imaging of the sample (\cref{fig:imaging}b) and thus precluding the need for free-space optics. Crucially, our sensor remains intact under the repetitive tip-sample contacts, highlighting the robust nature of the glued nanobeam-fiber assembly. 

With height control and navigation in place, we image the magnetic field of the strip by measuring the ESR frequency of the end-facet NVs (\cref{fig:imaging}c). To do so, we retract the tip by $\sim\SI{1}{\micron}$ after making tip-sample contact, measure the ESR spectrum and extract the end-facet NV ESR frequency using a double Lorentzian fit as described above (\cref{fig:misalign}c), with the sidewall ESR frequency fixed to \SI{2.729}{GHz} to increase the robustness of the fit. The extracted end-facet ESR frequencies (\cref{fig:imaging}c) encode the projection of the total magnetic field onto the NV axis. The expected near-uniform magnetic field within the ring and the magnetic field in the opposite direction outside the ring are clearly observed. 

A key parameter for determining the sensitivity of our sensor is the optical coupling efficiency between the nanobeam and the fiber. Specifically, the coupling efficiency $\eta$ of the end-facet NVs is the relevant parameter for our intended measurement regime. An estimation of this efficiency can be made via the optical saturation behavior of the measured photoluminescence \cite{Li2023ANanobeam}, combined with the estimated ratio between end-facet NVs and sidewall NVs from the ESR contrast ratio of the two dips (Methods). This results in $\eta = 2\%\text{-}26\%$, with the large uncertainty originating from the unknown conversion efficiency of implanted nitrogen ions to NV centers. This result is in reasonable agreement with that of fiber-coupled nanobeams with a known homogeneous NV density \cite{Li2023ANanobeam}. 

\section{Conclusion and Outlook}

We have demonstrated a robust, fiber-coupled scanning probe based on diamond nanobeams, utilizing NV centers for sub-micron resolution 2D magnetic imaging. Our fabrication approach, which incorporates angled ion implantation into the quasi-isotropic diamond etch workflow, deterministically positions NV centers at the end facet of the nanobeam. We attached these nanobeams to tapered fibers using optical glue to create a robust, transportable probe while optimizing the fiber-nanobeam interface by maximizing the photoluminescence detected through the fiber detection efficiency. Using our probe as a scanning magnetometer, we showed that we can isolate the signal of NV centers at the end facet even in the presence of the background signal from misaligned NV centers, and accurately extract the magnetic field variation at the sample surface. Finally, we demonstrate 2D imaging of the magnetic field generated by an omega-shaped current carrying wire under sub-micron spatial resolution. Combined with a method to image the sample surface through reflection-enhanced signal collection at probe-surface contact, we show that sample navigation and lift height control can be achieved without the need of free-space optical components, making our scanning probe ready for all-fiber operation in various environments.

Reducing the number of sidewall-implanted NVs would improve the signal-to-background ratio and thereby the magnetometry sensitivity. We anticipate this can be achieved by performing the angled ion implantation after the atomic layer deposition step (step iv in \cref{fig:fab}a): With a $\SI{20}{nm}$ layer of alumina uniformly covering the sample, one could set the implantation energy such that the nitrogen ions would be implanted into the diamond under the near-normal incidence at the end facet, but would get trapped in the alumina for the grazing-angle incidence at the beam sidewalls. 

The limiting factor of our spatial resolution is two-fold: First, the NV ensemble implanted homogeneously into the ($0.5\times\SI{0.5}{\micron\squared}$) beam end-facet limits the spatial resolution. Low-nanoscale imaging therefore requires moving to single NV regime. With the previous characterization of the negligible background contribution (\cref{fig:charac}), we already showed that single NV readout is possible under \SI{600}{\nm} + \SI{700}{\nm} long-pass filtering. Second, the reflection-based height control we used in this work does not allow feedback control of the tip-sample distance below $h\sim\SI{1}{\micron}$. To this end, we are working towards integrating our fiber probe in a tuning-fork based atomic force microscope (AFM) system \cite{Tung2008TappingFactor}, based on gluing the tip of the fiber to the tuning fork \cite{Gao2014DynamicSystem} or on keeping the fiber and the tuning fork in mechanical contact with a dedicated scan head design \cite{Tschirhart2021ImagingInsulator,Uri2020NanoscaleGraphene}.  

One important future application of our fiber-coupled probe is scanning magnetometry at low temperature, as free space optics is not required. As such, it could be an excellent platform for exploring the sensing capabilities of group-IV color centers, which have proven to be more stable against surface charge noise at low temperatures \cite{Rugar2021QuantumDiamond,DeSantis2021InvestigationDiamond,Rosenthal2023MicrowaveDiamond}. This potentially enables resonant optical addressing of target sensor defects in the beam. The nanobeam structure is also compatible with nanophotonic structuring, holding potential in further enhancement of excitation and readout efficiency of the color centers under resonant driving \cite{Pasini2023NonlinearWaveguide,Rugar2021QuantumDiamond}. With the above mentioned efforts, our fiber-coupled probe opens up new avenues for NV-based magnetic sensing in hard-to-reach environments.

\section{Methods}

\subsection{Fabrication of the diamond nanobeams}
We fabricate the diamond nanobeams using electronic grade single crystal diamond (Element Six). The fabrication recipe is based on ref.\cite{Pasini2023NonlinearWaveguide}, where details on process parameters can be found. Before the fabrication flow, the diamond substrate is sliced and polished into a $\SI{2}{\mm}\times\SI{2}{\mm}\times\SI{0.05}{\mm}$ chip (Almax EasyLab). After polishing, we clean the surface with fuming HNO$_3$ and relieve the surface damage induced by mechanical polishing with consecutive Ar/Cl and O$_2$ etches (Oxford Instruments Plasmalab 100). This removes a total of $6\text{-}\SI{8}{\micron}$ of diamond from the surface \cite{Simon2023BuildingWaves}.

We start the fabrication by depositing a \SI{200}{\nm} layer of Si$_3$N$_4$ hard mask with inductively-coupled plasma enhanced chemical vapour deposition (ICPECVD) (Oxford Instruments PlasmaPro 100). Then we spin-coat a $\sim\SI{400}{\nm}$ layer of e-beam resist (AR-P 6200-13) and another $\sim\SI{30}{\nm}$ layer of conductive Electra-92, and write the pattern with e-beam lithography (Raith EBPG5200). The pattern is transferred to the hard mask with an anisotropic CHF$_3$/O$_2$ etch (AMS 100 I-speeder). After cleaning the residual e-beam resist with dimethylformamide (DMF) and
piranha solution (3:1 mixed 96\% H$_2$SO$_4$ and 31\% H$_2$O$_2$), we anisotropically etch $\sim\SI{600}{\nm}$ into the diamond with O$_2$ RIE (Oxford Instruments Plasmalab 100) to create the sidewalls of the nanobeams.

We then implant $^{14}$N ions into the diamond sidewalls as described in the main text (INNOViON Corp). After implantation, we do another piranha cleaning to remove contamination on the surface, and deposit $\sim\SI{20}{\nm}$ of Al$_2$O$_3$ with atomic layer deposition (ALD) (Oxford Instruments FlexAL). We then do an anisotropic BCl$_3$/Cl$_2$ RIE (Oxford Instruments Plasmalab 100) to etch away the Al$_2$O$_3$ on the top surfaces, so that only the sidewalls of the diamond nanobeams remain protected while the diamond surfaces around are exposed to O$_2$ plasma in subsequent steps. We do a second anisotropic O$_2$ RIE which etches another $\sim\SI{250}{\nm}$ into the diamond. This is to eliminate the potential effect of the damaged diamond surface after the ion implantation to the undercut quality. We then proceed with the quasi-isotropic O$_2$ RIE to undercut the beams and remove the masks with 40\% HF after the beams are fully released. The fabricated diamond chip is then vacuum annealed at \SI{800}{\celsius} for 2 hours to create the NV centers.

\subsection{Setup for current measurements}

The gold strips for current measurements are patterned on top of a silicon substrate via evaporation (\SI{5}{\nm} titanium + \SI{100}{\nm} gold, Temescal FC-2000) and lift-off. The DC current is applied using a function generator (Tektronix AFG1062) and read out by a digital multimeter wired in series with the sample. We apply the RF signal for driving the NV centers through the same strip, using a bias-T (Mini-Circuits ZFBT-6GW-FT+) to combine the RF and DC signal. A complete schematic of the measurement setup can be found in SI.

\subsection{Vector magnetometry with NV centers}

Under an external magnetic field $\mathbf{B}$, the Hamiltonian of an NV center is written as
\begin{equation}
    H = DS_z^2 + \gamma\mathbf{B}\cdot\mathbf{S}
\end{equation}
in which $D=\SI{2.87}{GHz}$ is the zero-field splitting of the NV center, $\gamma=\SI{28}{\GHz\per\tesla}$ is the gyromagnatic ratio and $\mathbf{S}=(S_x,S_y,S_z)$ are the $3\times3$ Pauli matrices for spin-1. The upper and lower ESR frequencies $f_u$ and $f_l$ are then the differences between the eigenfrequencies of the Hamiltonian. By diagonalizing the Hamiltonian, one can determine from the measured ESR frequencies both the amplitude of the external field $B$, and its angle with respect to the NV orientation $\theta$:
\begin{equation}
    B = \frac{1}{\sqrt{3}\gamma}\sqrt{f_u^2 + f_l^2 - f_uf_l + D^2};
\end{equation}
\begin{equation}
    \cos^2{\theta} = \frac{-(f_u+f_l)^3+3(f_u^3+f_l^3)+2D^3}{27D(\gamma B)^2} + \frac{1}{3}.
\end{equation}

These relations allow us to align the external field $B_0$ with the NV orientation, and precisely determine the value of $B_0$ through the measured ESR spectrum during calibration. In the main text, we also used the inverse of these relations to calculate the expected ESR frequencies of a given magnetic field distribution from the DC current, and used this calculation to fit the experimentally measured ESR frequencies in \cref{fig:misalign}(e).

\subsection{Estimation of the coupling efficiency for the end-facet NVs}

The fraction of end-facet NV contribution in the total measured PL can be estimated from the contrast ratio extracted from \cref{fig:misalign}c left (where the dips are most split): 
\begin{equation}
    \beta_\mathrm{E} = \frac{I_\mathrm{E}}{I_\mathrm{tot}} = \frac{C_\mathrm{E}}{C_\mathrm{E} + C_\mathrm{S}} = 0.43(7)
\end{equation}
where the footnotes E and S stand for the end-facet and sidewall contribution respectively. We then scale the \SI{600}{nm} filtered saturation curve in \cref{fig:charac}c with $\beta_\mathrm{E}$ to extract the saturation intensity of the end-facet NVs
\begin{equation}
    I_\mathrm{sat,E} = \beta_\mathrm{E}I_\mathrm{sat,600 nm} = (5.6\pm1.1)\times10^5 \mathrm{s}^{-1},
\end{equation}
from which the total collection efficiency of the end-facet NVs can be estimated as the ratio of the detected saturation intensity and the theoretically expected saturation emission rate:
\begin{equation}
    \eta_\mathrm{E} = \frac{I_\mathrm{sat,E}}{N_\mathrm{E}\Gamma},
\end{equation}
where $N_\mathrm{E}$ denotes the number of end-facet NVs, and $\Gamma = 1/(\SI{13}{ns})$ \cite{Manson2006Nitrogen-vacancyDynamics} is the expected saturation photon emission rate of a single NV. In our measurement setup, $\eta_\mathrm{E}$ can be expressed as the product of the fiber-nanobeam coupling efficiency $\eta_\mathrm{E,f}$, ND filtering $\eta_\mathrm{ND} = 1\times10^{-3}$ and the total detection efficiency of the optical path $\eta_\mathrm{D} = 0.14(2)$ (including the loss on optical elements and detection efficiency of the APD).

The major uncertainty in estimating the fiber-nanobeam coupling efficiency $\eta_\mathrm{E,f}$ lies in the NV number $N_\mathrm{E}$, due to the large uncertainty in the conversion efficiency of the implanted nitrogen ions to NV centers. Taking the data from ref.\cite{Pezzagna2010CreationDiamond} as a reference and further considering the possible deviation from this dataset caused by the presence of nanostructures, we assume the conversion efficiency to be between 1\%-10\%. This results in $\eta_\mathrm{E,f} = 2\%\text{-}26\%$.

\section{Author Imformation}

\subsection{Funding Sources}
This work is supported by the Dutch Science Council (NWO) through the NWA grant 1160.18.208, the NGF Quantum Technology grant 1582.22.038 and the Kavli Institute of Nanoscience Delft.

\subsection{Author Contributions}
Y.L., R.N. and T.v.d.S. conceived the experiments. Y.L. fabricated the diamond nanobeam samples and performed the magnetometry measurements. Y.L. and G.W. developed the gluing workflow and prepared the fiber-coupled nanobeam probes. Y.L. and T.v.d.S. analyzed the results, and wrote the manuscript with contributions from all coauthors.

 \subsection{Notes}
The authors declare no competing financial interest.

 \subsection{Data availability}
 All data plotted in the figures are this work are available at zenodo.org with identifier 10.5281/zenodo.12724082. Additional data related to this paper are available upon request.

\begin{acknowledgement}
The authors thank N. Codreanu for the help with the diamond nanofabrication process, and B.G. Simon for helping with diamond preparation and ion implantation.

\end{acknowledgement}

\begin{suppinfo}
Setup for nanobeam manipulation and scanning measurements; Analysis on background photoluminescence and the effect of filtering wavelengths; Time trace of measured photoluminescence upon nanobeam-sample contact; Fit results for the 2D scan.

\end{suppinfo}

\pagebreak
\bibliography{references}

\providecommand{\latin}[1]{#1}
\makeatletter
\providecommand{\doi}
  {\begingroup\let\do\@makeother\dospecials
  \catcode`\{=1 \catcode`\}=2 \doi@aux}
\providecommand{\doi@aux}[1]{\endgroup\texttt{#1}}
\makeatother
\providecommand*\mcitethebibliography{\thebibliography}
\csname @ifundefined\endcsname{endmcitethebibliography}  {\let\endmcitethebibliography\endthebibliography}{}
\begin{mcitethebibliography}{33}
\providecommand*\natexlab[1]{#1}
\providecommand*\mciteSetBstSublistMode[1]{}
\providecommand*\mciteSetBstMaxWidthForm[2]{}
\providecommand*\mciteBstWouldAddEndPuncttrue
  {\def\EndOfBibitem{\unskip.}}
\providecommand*\mciteBstWouldAddEndPunctfalse
  {\let\EndOfBibitem\relax}
\providecommand*\mciteSetBstMidEndSepPunct[3]{}
\providecommand*\mciteSetBstSublistLabelBeginEnd[3]{}
\providecommand*\EndOfBibitem{}
\mciteSetBstSublistMode{f}
\mciteSetBstMaxWidthForm{subitem}{(\alph{mcitesubitemcount})}
\mciteSetBstSublistLabelBeginEnd
  {\mcitemaxwidthsubitemform\space}
  {\relax}
  {\relax}

\bibitem[Balasubramanian \latin{et~al.}(2008)Balasubramanian, Chan, Kolesov, Al-Hmoud, Tisler, Shin, Kim, Wojcik, Hemmer, Krueger, Hanke, Leitenstorfer, Bratschitsch, Jelezko, and Wrachtrup]{Balasubramanian2008NanoscaleConditions}
Balasubramanian,~G.; Chan,~I.~Y.; Kolesov,~R.; Al-Hmoud,~M.; Tisler,~J.; Shin,~C.; Kim,~C.; Wojcik,~A.; Hemmer,~P.~R.; Krueger,~A.; Hanke,~T.; Leitenstorfer,~A.; Bratschitsch,~R.; Jelezko,~F.; Wrachtrup,~J. {Nanoscale imaging magnetometry with diamond spins under ambient conditions}. \emph{Nature} \textbf{2008}, \emph{455}, 648--651\relax
\mciteBstWouldAddEndPuncttrue
\mciteSetBstMidEndSepPunct{\mcitedefaultmidpunct}
{\mcitedefaultendpunct}{\mcitedefaultseppunct}\relax
\EndOfBibitem
\bibitem[Maze \latin{et~al.}(2008)Maze, Stanwix, Hodges, Hong, Taylor, Cappellaro, Jiang, Dutt, Togan, Zibrov, Yacoby, Walsworth, and Lukin]{Maze2008NanoscaleDiamond}
Maze,~J.~R.; Stanwix,~P.~L.; Hodges,~J.~S.; Hong,~S.; Taylor,~J.~M.; Cappellaro,~P.; Jiang,~L.; Dutt,~M. V.~G.; Togan,~E.; Zibrov,~A.~S.; Yacoby,~A.; Walsworth,~R.~L.; Lukin,~M.~D. {Nanoscale magnetic sensing with an individual electronic spin in diamond}. \emph{Nature} \textbf{2008}, \emph{455}, 644--647\relax
\mciteBstWouldAddEndPuncttrue
\mciteSetBstMidEndSepPunct{\mcitedefaultmidpunct}
{\mcitedefaultendpunct}{\mcitedefaultseppunct}\relax
\EndOfBibitem
\bibitem[Maletinsky \latin{et~al.}(2012)Maletinsky, Hong, Grinolds, Hausmann, Lukin, Walsworth, Loncar, and Yacoby]{Maletinsky2012ACentres}
Maletinsky,~P.; Hong,~S.; Grinolds,~M.~S.; Hausmann,~B.; Lukin,~M.~D.; Walsworth,~R.~L.; Loncar,~M.; Yacoby,~A. {A robust scanning diamond sensor for nanoscale imaging with single nitrogen-vacancy centres}. \emph{Nature Nanotechnology} \textbf{2012}, \emph{7}, 320--324\relax
\mciteBstWouldAddEndPuncttrue
\mciteSetBstMidEndSepPunct{\mcitedefaultmidpunct}
{\mcitedefaultendpunct}{\mcitedefaultseppunct}\relax
\EndOfBibitem
\bibitem[Casola \latin{et~al.}(2018)Casola, Van Der~Sar, and Yacoby]{Casola2018ProbingDiamond}
Casola,~F.; Van Der~Sar,~T.; Yacoby,~A. {Probing condensed matter physics with magnetometry based on nitrogen-vacancy centres in diamond}. \emph{Nature Reviews Materials} \textbf{2018}, \emph{3}\relax
\mciteBstWouldAddEndPuncttrue
\mciteSetBstMidEndSepPunct{\mcitedefaultmidpunct}
{\mcitedefaultendpunct}{\mcitedefaultseppunct}\relax
\EndOfBibitem
\bibitem[Schirhagl \latin{et~al.}(2014)Schirhagl, Chang, Loretz, and Degen]{Schirhagl2014Nitrogen-vacancyBiology}
Schirhagl,~R.; Chang,~K.; Loretz,~M.; Degen,~C.~L. {Nitrogen-vacancy centers in diamond: Nanoscale sensors for physics and biology}. \emph{Annual Review of Physical Chemistry} \textbf{2014}, \emph{65}, 83--105\relax
\mciteBstWouldAddEndPuncttrue
\mciteSetBstMidEndSepPunct{\mcitedefaultmidpunct}
{\mcitedefaultendpunct}{\mcitedefaultseppunct}\relax
\EndOfBibitem
\bibitem[Le~Sage \latin{et~al.}(2013)Le~Sage, Arai, Glenn, Devience, Pham, Rahn-Lee, Lukin, Yacoby, Komeili, and Walsworth]{LeSage2013OpticalCells}
Le~Sage,~D.; Arai,~K.; Glenn,~D.~R.; Devience,~S.~J.; Pham,~L.~M.; Rahn-Lee,~L.; Lukin,~M.~D.; Yacoby,~A.; Komeili,~A.; Walsworth,~R.~L. {Optical magnetic imaging of living cells}. \emph{Nature} \textbf{2013}, \emph{496}, 486--489\relax
\mciteBstWouldAddEndPuncttrue
\mciteSetBstMidEndSepPunct{\mcitedefaultmidpunct}
{\mcitedefaultendpunct}{\mcitedefaultseppunct}\relax
\EndOfBibitem
\bibitem[Steinert \latin{et~al.}(2013)Steinert, Ziem, Hall, Zappe, Schweikert, G{\"{o}}tz, Aird, Balasubramanian, Hollenberg, and Wrachtrup]{Steinert2013MagneticResolution}
Steinert,~S.; Ziem,~F.; Hall,~L.~T.; Zappe,~A.; Schweikert,~M.; G{\"{o}}tz,~N.; Aird,~A.; Balasubramanian,~G.; Hollenberg,~L.; Wrachtrup,~J. {Magnetic spin imaging under ambient conditions with sub-cellular resolution}. \emph{Nature Communications} \textbf{2013}, \emph{4}\relax
\mciteBstWouldAddEndPuncttrue
\mciteSetBstMidEndSepPunct{\mcitedefaultmidpunct}
{\mcitedefaultendpunct}{\mcitedefaultseppunct}\relax
\EndOfBibitem
\bibitem[de~Groot \latin{et~al.}(2021)de~Groot, Fabian, B{\'{e}}guin, Kosters, Cort{\'{e}}s-Ortu{\~{n}}o, Fu, Jansen, Harrison, van Leeuwen, and Barnhoorn]{deGroot2021MicromagneticRock-Magnetism}
de~Groot,~L.~V.; Fabian,~K.; B{\'{e}}guin,~A.; Kosters,~M.~E.; Cort{\'{e}}s-Ortu{\~{n}}o,~D.; Fu,~R.~R.; Jansen,~C.~M.; Harrison,~R.~J.; van Leeuwen,~T.; Barnhoorn,~A. {Micromagnetic Tomography for Paleomagnetism and Rock-Magnetism}. \emph{Journal of Geophysical Research: Solid Earth} \textbf{2021}, \emph{126}\relax
\mciteBstWouldAddEndPuncttrue
\mciteSetBstMidEndSepPunct{\mcitedefaultmidpunct}
{\mcitedefaultendpunct}{\mcitedefaultseppunct}\relax
\EndOfBibitem
\bibitem[Gross \latin{et~al.}(2017)Gross, Akhtar, Garcia, Mart{\'{i}}nez, Chouaieb, Garcia, Carr{\'{e}}t{\'{e}}ro, Barth{\'{e}}l{\'{e}}my, Appel, Maletinsky, Kim, Chauleau, Jaouen, Viret, Bibes, Fusil, and Jacques]{Gross2017Real-spaceMagnetometer}
Gross,~I. \latin{et~al.}  {Real-space imaging of non-collinear antiferromagnetic order with a single-spin magnetometer}. \emph{Nature} \textbf{2017}, \emph{549}, 252--256\relax
\mciteBstWouldAddEndPuncttrue
\mciteSetBstMidEndSepPunct{\mcitedefaultmidpunct}
{\mcitedefaultendpunct}{\mcitedefaultseppunct}\relax
\EndOfBibitem
\bibitem[Thiel \latin{et~al.}(2019)Thiel, Wang, Tschudin, Rohner, Guti{\'{e}}rrez-Lezama, Ubrig, Gibertini, Giannini, Morpurgo, and Maletinsky]{Thiel2019ProbingMicroscopy}
Thiel,~L.; Wang,~Z.; Tschudin,~M.~A.; Rohner,~D.; Guti{\'{e}}rrez-Lezama,~I.; Ubrig,~N.; Gibertini,~M.; Giannini,~E.; Morpurgo,~A.~F.; Maletinsky,~P. {Probing magnetism in 2D materials at the nanoscale with single-spin microscopy}. \emph{Science} \textbf{2019}, \emph{364}, 973--976\relax
\mciteBstWouldAddEndPuncttrue
\mciteSetBstMidEndSepPunct{\mcitedefaultmidpunct}
{\mcitedefaultendpunct}{\mcitedefaultseppunct}\relax
\EndOfBibitem
\bibitem[Sun \latin{et~al.}(2021)Sun, Song, Anderson, Brunner, F{\"{o}}rster, Shalomayeva, Taniguchi, Watanabe, Gr{\"{a}}fe, St{\"{o}}hr, Xu, and Wrachtrup]{Sun2021MagneticImaging}
Sun,~Q.-C.; Song,~T.; Anderson,~E.; Brunner,~A.; F{\"{o}}rster,~J.; Shalomayeva,~T.; Taniguchi,~T.; Watanabe,~K.; Gr{\"{a}}fe,~J.; St{\"{o}}hr,~R.; Xu,~X.; Wrachtrup,~J. {Magnetic domains and domain wall pinning in atomically thin CrBr3 revealed by nanoscale imaging}. \emph{Nature Communications} \textbf{2021}, \emph{12}, 1989\relax
\mciteBstWouldAddEndPuncttrue
\mciteSetBstMidEndSepPunct{\mcitedefaultmidpunct}
{\mcitedefaultendpunct}{\mcitedefaultseppunct}\relax
\EndOfBibitem
\bibitem[Simon \latin{et~al.}(2021)Simon, Kurdi, La, Bertelli, Carmiggelt, Ruf, De~Jong, Van Den~Berg, Katan, and Van Der~Sar]{Simon2021DirectionalWaves}
Simon,~B.~G.; Kurdi,~S.; La,~H.; Bertelli,~I.; Carmiggelt,~J.~J.; Ruf,~M.; De~Jong,~N.; Van Den~Berg,~H.; Katan,~A.~J.; Van Der~Sar,~T. {Directional Excitation of a High-Density Magnon Gas Using Coherently Driven Spin Waves}. \emph{Nano Letters} \textbf{2021}, \emph{21}, 8213--8219\relax
\mciteBstWouldAddEndPuncttrue
\mciteSetBstMidEndSepPunct{\mcitedefaultmidpunct}
{\mcitedefaultendpunct}{\mcitedefaultseppunct}\relax
\EndOfBibitem
\bibitem[Fedotov \latin{et~al.}(2014)Fedotov, Doronina-Amitonova, Sidorov-Biryukov, Safronov, Blakley, Levchenko, Zibrov, Fedotov, Kilin, Scully, Velichansky, and Zheltikov]{Fedotov2014Fiber-opticImaging}
Fedotov,~I.~V.; Doronina-Amitonova,~L.~V.; Sidorov-Biryukov,~D.~A.; Safronov,~N.~A.; Blakley,~S.; Levchenko,~A.~O.; Zibrov,~S.~A.; Fedotov,~A.~B.; Kilin,~S.~Y.; Scully,~M.~O.; Velichansky,~V.~L.; Zheltikov,~A.~M. {Fiber-optic magnetic-field imaging}. \emph{Optics Letters} \textbf{2014}, \emph{39}, 6954\relax
\mciteBstWouldAddEndPuncttrue
\mciteSetBstMidEndSepPunct{\mcitedefaultmidpunct}
{\mcitedefaultendpunct}{\mcitedefaultseppunct}\relax
\EndOfBibitem
\bibitem[Chatzidrosos \latin{et~al.}(2021)Chatzidrosos, Rebeirro, Zheng, Omar, Brenneis, St{\"{u}}rner, Fuchs, Buck, R{\"{o}}lver, Schneemann, Bl{\"{u}}mler, Budker, and Wickenbrock]{Chatzidrosos2021FiberizedMagnetometers}
Chatzidrosos,~G.; Rebeirro,~J.~S.; Zheng,~H.; Omar,~M.; Brenneis,~A.; St{\"{u}}rner,~F.~M.; Fuchs,~T.; Buck,~T.; R{\"{o}}lver,~R.; Schneemann,~T.; Bl{\"{u}}mler,~P.; Budker,~D.; Wickenbrock,~A. {Fiberized Diamond-Based Vector Magnetometers}. \emph{Frontiers in Photonics} \textbf{2021}, \emph{2}\relax
\mciteBstWouldAddEndPuncttrue
\mciteSetBstMidEndSepPunct{\mcitedefaultmidpunct}
{\mcitedefaultendpunct}{\mcitedefaultseppunct}\relax
\EndOfBibitem
\bibitem[Dix \latin{et~al.}(2024)Dix, L{\"{o}}nard, Barbosa, Gutsche, Witzenrath, and Widera]{Dix2024ACenters}
Dix,~S.; L{\"{o}}nard,~D.; Barbosa,~I.~C.; Gutsche,~J.; Witzenrath,~J.; Widera,~A. {A miniaturized magnetic field sensor based on nitrogen-vacancy centers}. \emph{arXiv:2402.19372} \textbf{2024}, \relax
\mciteBstWouldAddEndPunctfalse
\mciteSetBstMidEndSepPunct{\mcitedefaultmidpunct}
{}{\mcitedefaultseppunct}\relax
\EndOfBibitem
\bibitem[Parker \latin{et~al.}(2024)Parker, Arjona~Mart{\'{i}}nez, Chen, Stramma, Harris, Michaels, Trusheim, Hayhurst~Appel, Purser, Roth, Englund, and Atat{\"{u}}re]{Parker2024ARegister}
Parker,~R.~A.; Arjona~Mart{\'{i}}nez,~J.; Chen,~K.~C.; Stramma,~A.~M.; Harris,~I.~B.; Michaels,~C.~P.; Trusheim,~M.~E.; Hayhurst~Appel,~M.; Purser,~C.~M.; Roth,~W.~G.; Englund,~D.; Atat{\"{u}}re,~M. {A diamond nanophotonic interface with an optically accessible deterministic electronuclear spin register}. \emph{Nature Photonics} \textbf{2024}, \emph{18}, 156--161\relax
\mciteBstWouldAddEndPuncttrue
\mciteSetBstMidEndSepPunct{\mcitedefaultmidpunct}
{\mcitedefaultendpunct}{\mcitedefaultseppunct}\relax
\EndOfBibitem
\bibitem[Pasini \latin{et~al.}(2023)Pasini, Codreanu, Turan, Moral, Primavera, De~Santis, Beukers, Brevoord, Waas, Borregaard, and Hanson]{Pasini2023NonlinearWaveguide}
Pasini,~M.; Codreanu,~N.; Turan,~T.; Moral,~A.~R.; Primavera,~C.~F.; De~Santis,~L.; Beukers,~H. K.~C.; Brevoord,~J.~M.; Waas,~C.; Borregaard,~J.; Hanson,~R. {Nonlinear Quantum Photonics with a Tin-Vacancy Center Coupled to a One-Dimensional Diamond Waveguide}. \emph{arXiv:2311.12927} \textbf{2023}, \relax
\mciteBstWouldAddEndPunctfalse
\mciteSetBstMidEndSepPunct{\mcitedefaultmidpunct}
{}{\mcitedefaultseppunct}\relax
\EndOfBibitem
\bibitem[Ziegler(2013)]{Ziegler2013SRIM-2013}
Ziegler,~J. {SRIM-2013}. 2013; \url{http://www.srim.org/}\relax
\mciteBstWouldAddEndPuncttrue
\mciteSetBstMidEndSepPunct{\mcitedefaultmidpunct}
{\mcitedefaultendpunct}{\mcitedefaultseppunct}\relax
\EndOfBibitem
\bibitem[Mouradian \latin{et~al.}(2017)Mouradian, Wan, Schr{\"{o}}der, and Englund]{Mouradian2017RectangularDiamond}
Mouradian,~S.; Wan,~N.~H.; Schr{\"{o}}der,~T.; Englund,~D. {Rectangular photonic crystal nanobeam cavities in bulk diamond}. \emph{Applied Physics Letters} \textbf{2017}, \emph{111}\relax
\mciteBstWouldAddEndPuncttrue
\mciteSetBstMidEndSepPunct{\mcitedefaultmidpunct}
{\mcitedefaultendpunct}{\mcitedefaultseppunct}\relax
\EndOfBibitem
\bibitem[Li \latin{et~al.}(2023)Li, Gerritsma, Kurdi, Codreanu, Gr{\"{o}}blacher, Hanson, Norte, and van~der Sar]{Li2023ANanobeam}
Li,~Y.; Gerritsma,~F.~A.; Kurdi,~S.; Codreanu,~N.; Gr{\"{o}}blacher,~S.; Hanson,~R.; Norte,~R.; van~der Sar,~T. {A Fiber-Coupled Scanning Magnetometer with Nitrogen-Vacancy Spins in a Diamond Nanobeam}. \emph{ACS Photonics} \textbf{2023}, \emph{10}, 1859--1865\relax
\mciteBstWouldAddEndPuncttrue
\mciteSetBstMidEndSepPunct{\mcitedefaultmidpunct}
{\mcitedefaultendpunct}{\mcitedefaultseppunct}\relax
\EndOfBibitem
\bibitem[Dr{\'{e}}au \latin{et~al.}(2011)Dr{\'{e}}au, Lesik, Rondin, Spinicelli, Arcizet, Roch, and Jacques]{Dreau2011AvoidingSensitivity}
Dr{\'{e}}au,~A.; Lesik,~M.; Rondin,~L.; Spinicelli,~P.; Arcizet,~O.; Roch,~J.~F.; Jacques,~V. {Avoiding power broadening in optically detected magnetic resonance of single NV defects for enhanced dc magnetic field sensitivity}. \emph{Physical Review B - Condensed Matter and Materials Physics} \textbf{2011}, \emph{84}\relax
\mciteBstWouldAddEndPuncttrue
\mciteSetBstMidEndSepPunct{\mcitedefaultmidpunct}
{\mcitedefaultendpunct}{\mcitedefaultseppunct}\relax
\EndOfBibitem
\bibitem[Zheng \latin{et~al.}(2022)Zheng, Bian, Chen, Shen, Zhang, St{\"{o}}hr, Denisenko, Wrachtrup, Yang, and Jiang]{Zheng2022CoherenceBath}
Zheng,~W.; Bian,~K.; Chen,~X.; Shen,~Y.; Zhang,~S.; St{\"{o}}hr,~R.; Denisenko,~A.; Wrachtrup,~J.; Yang,~S.; Jiang,~Y. {Coherence enhancement of solid-state qubits by local manipulation of the electron spin bath}. \emph{Nature Physics} \textbf{2022}, \emph{18}, 1317--1323\relax
\mciteBstWouldAddEndPuncttrue
\mciteSetBstMidEndSepPunct{\mcitedefaultmidpunct}
{\mcitedefaultendpunct}{\mcitedefaultseppunct}\relax
\EndOfBibitem
\bibitem[Tung \latin{et~al.}(2008)Tung, Chizhik, Chikunov, and Hoai]{Tung2008TappingFactor}
Tung,~V.~T.; Chizhik,~S.~A.; Chikunov,~V.~V.; Hoai,~T.~X. {Tapping and shear-mode atomic force microscopy using a quartz tuning fork with high quality factor}. Twelfth International Workshop on Nanodesign Technology and Computer Simulations. 2008; p 73770L\relax
\mciteBstWouldAddEndPuncttrue
\mciteSetBstMidEndSepPunct{\mcitedefaultmidpunct}
{\mcitedefaultendpunct}{\mcitedefaultseppunct}\relax
\EndOfBibitem
\bibitem[Gao \latin{et~al.}(2014)Gao, Li, Wang, and Fu]{Gao2014DynamicSystem}
Gao,~F.; Li,~X.; Wang,~J.; Fu,~Y. {Dynamic behavior of tuning fork shear-force structures in a SNOM system}. \emph{Ultramicroscopy} \textbf{2014}, \emph{142}, 10--23\relax
\mciteBstWouldAddEndPuncttrue
\mciteSetBstMidEndSepPunct{\mcitedefaultmidpunct}
{\mcitedefaultendpunct}{\mcitedefaultseppunct}\relax
\EndOfBibitem
\bibitem[Tschirhart \latin{et~al.}(2021)Tschirhart, Serlin, Polshyn, Shragai, Xia, Zhu, Zhang, Watanabe, Taniguchi, Huber, and Young]{Tschirhart2021ImagingInsulator}
Tschirhart,~C.~L.; Serlin,~M.; Polshyn,~H.; Shragai,~A.; Xia,~Z.; Zhu,~J.; Zhang,~Y.; Watanabe,~K.; Taniguchi,~T.; Huber,~M.~E.; Young,~A.~F. {Imaging orbital ferromagnetism in a moir{\'{e}} Chern insulator}. \emph{Science} \textbf{2021}, \emph{372}, 1323--1327\relax
\mciteBstWouldAddEndPuncttrue
\mciteSetBstMidEndSepPunct{\mcitedefaultmidpunct}
{\mcitedefaultendpunct}{\mcitedefaultseppunct}\relax
\EndOfBibitem
\bibitem[Uri \latin{et~al.}(2020)Uri, Kim, Bagani, Lewandowski, Grover, Auerbach, Lachman, Myasoedov, Taniguchi, Watanabe, Smet, and Zeldov]{Uri2020NanoscaleGraphene}
Uri,~A.; Kim,~Y.; Bagani,~K.; Lewandowski,~C.~K.; Grover,~S.; Auerbach,~N.; Lachman,~E.~O.; Myasoedov,~Y.; Taniguchi,~T.; Watanabe,~K.; Smet,~J.; Zeldov,~E. {Nanoscale imaging of equilibrium quantum Hall edge currents and of the magnetic monopole response in graphene}. \emph{Nature Physics} \textbf{2020}, \emph{16}, 164--170\relax
\mciteBstWouldAddEndPuncttrue
\mciteSetBstMidEndSepPunct{\mcitedefaultmidpunct}
{\mcitedefaultendpunct}{\mcitedefaultseppunct}\relax
\EndOfBibitem
\bibitem[Rugar \latin{et~al.}(2021)Rugar, Aghaeimeibodi, Riedel, Dory, Lu, McQuade, Shen, Melosh, and Vu{\v{c}}kovi{\'{c}}]{Rugar2021QuantumDiamond}
Rugar,~A.~E.; Aghaeimeibodi,~S.; Riedel,~D.; Dory,~C.; Lu,~H.; McQuade,~P.~J.; Shen,~Z.-X.; Melosh,~N.~A.; Vu{\v{c}}kovi{\'{c}},~J. {Quantum Photonic Interface for Tin-Vacancy Centers in Diamond}. \emph{Physical Review X} \textbf{2021}, \emph{11}, 031021\relax
\mciteBstWouldAddEndPuncttrue
\mciteSetBstMidEndSepPunct{\mcitedefaultmidpunct}
{\mcitedefaultendpunct}{\mcitedefaultseppunct}\relax
\EndOfBibitem
\bibitem[De~Santis \latin{et~al.}(2021)De~Santis, Trusheim, Chen, and Englund]{DeSantis2021InvestigationDiamond}
De~Santis,~L.; Trusheim,~M.~E.; Chen,~K.~C.; Englund,~D.~R. {Investigation of the Stark Effect on a Centrosymmetric Quantum Emitter in Diamond}. \emph{Physical Review Letters} \textbf{2021}, \emph{127}, 147402\relax
\mciteBstWouldAddEndPuncttrue
\mciteSetBstMidEndSepPunct{\mcitedefaultmidpunct}
{\mcitedefaultendpunct}{\mcitedefaultseppunct}\relax
\EndOfBibitem
\bibitem[Rosenthal \latin{et~al.}(2023)Rosenthal, Anderson, Kleidermacher, Stein, Lee, Grzesik, Scuri, Rugar, Riedel, Aghaeimeibodi, Ahn, Van~Gasse, and Vu{\v{c}}kovi{\'{c}}]{Rosenthal2023MicrowaveDiamond}
Rosenthal,~E.~I.; Anderson,~C.~P.; Kleidermacher,~H.~C.; Stein,~A.~J.; Lee,~H.; Grzesik,~J.; Scuri,~G.; Rugar,~A.~E.; Riedel,~D.; Aghaeimeibodi,~S.; Ahn,~G.~H.; Van~Gasse,~K.; Vu{\v{c}}kovi{\'{c}},~J. {Microwave Spin Control of a Tin-Vacancy Qubit in Diamond}. \emph{Physical Review X} \textbf{2023}, \emph{13}, 031022\relax
\mciteBstWouldAddEndPuncttrue
\mciteSetBstMidEndSepPunct{\mcitedefaultmidpunct}
{\mcitedefaultendpunct}{\mcitedefaultseppunct}\relax
\EndOfBibitem
\bibitem[Simon(2023)]{Simon2023BuildingWaves}
Simon,~B. {Building a platform for magnetic imaging of spin waves}. Ph.D.\ thesis, TU Delft, 2023\relax
\mciteBstWouldAddEndPuncttrue
\mciteSetBstMidEndSepPunct{\mcitedefaultmidpunct}
{\mcitedefaultendpunct}{\mcitedefaultseppunct}\relax
\EndOfBibitem
\bibitem[Manson \latin{et~al.}(2006)Manson, Harrison, and Sellars]{Manson2006Nitrogen-vacancyDynamics}
Manson,~N.~B.; Harrison,~J.~P.; Sellars,~M.~J. {Nitrogen-vacancy center in diamond: Model of the electronic structure and associated dynamics}. \emph{Physical Review B - Condensed Matter and Materials Physics} \textbf{2006}, \emph{74}\relax
\mciteBstWouldAddEndPuncttrue
\mciteSetBstMidEndSepPunct{\mcitedefaultmidpunct}
{\mcitedefaultendpunct}{\mcitedefaultseppunct}\relax
\EndOfBibitem
\bibitem[Pezzagna \latin{et~al.}(2010)Pezzagna, Naydenov, Jelezko, Wrachtrup, and Meijer]{Pezzagna2010CreationDiamond}
Pezzagna,~S.; Naydenov,~B.; Jelezko,~F.; Wrachtrup,~J.; Meijer,~J. {Creation efficiency of nitrogen-vacancy centres in diamond}. \emph{New Journal of Physics} \textbf{2010}, \emph{12}\relax
\mciteBstWouldAddEndPuncttrue
\mciteSetBstMidEndSepPunct{\mcitedefaultmidpunct}
{\mcitedefaultendpunct}{\mcitedefaultseppunct}\relax
\EndOfBibitem
\end{mcitethebibliography}


\providecommand{\latin}[1]{#1}
\makeatletter
\providecommand{\doi}
  {\begingroup\let\do\@makeother\dospecials
  \catcode`\{=1 \catcode`\}=2 \doi@aux}
\providecommand{\doi@aux}[1]{\endgroup\texttt{#1}}
\makeatother
\providecommand*\mcitethebibliography{\thebibliography}
\csname @ifundefined\endcsname{endmcitethebibliography}  {\let\endmcitethebibliography\endthebibliography}{}
\begin{mcitethebibliography}{4}
\providecommand*\natexlab[1]{#1}
\providecommand*\mciteSetBstSublistMode[1]{}
\providecommand*\mciteSetBstMaxWidthForm[2]{}
\providecommand*\mciteBstWouldAddEndPuncttrue
  {\def\EndOfBibitem{\unskip.}}
\providecommand*\mciteBstWouldAddEndPunctfalse
  {\let\EndOfBibitem\relax}
\providecommand*\mciteSetBstMidEndSepPunct[3]{}
\providecommand*\mciteSetBstSublistLabelBeginEnd[3]{}
\providecommand*\EndOfBibitem{}
\mciteSetBstSublistMode{f}
\mciteSetBstMaxWidthForm{subitem}{(\alph{mcitesubitemcount})}
\mciteSetBstSublistLabelBeginEnd
  {\mcitemaxwidthsubitemform\space}
  {\relax}
  {\relax}

\bibitem[Savvin \latin{et~al.}(2021)Savvin, Dormidonov, Smetanina, Mitrokhin, Lipatov, Genin, Potanin, Yelisseyev, and Vins]{Savvin2021NVLaser}
Savvin,~A.; Dormidonov,~A.; Smetanina,~E.; Mitrokhin,~V.; Lipatov,~E.; Genin,~D.; Potanin,~S.; Yelisseyev,~A.; Vins,~V. {NV– diamond laser}. \emph{Nature Communications} \textbf{2021}, \emph{12}, 7118\relax
\mciteBstWouldAddEndPuncttrue
\mciteSetBstMidEndSepPunct{\mcitedefaultmidpunct}
{\mcitedefaultendpunct}{\mcitedefaultseppunct}\relax
\EndOfBibitem
\bibitem[Cardoso~Barbosa \latin{et~al.}(2023)Cardoso~Barbosa, Gutsche, and Widera]{CardosoBarbosa2023ImpactRelaxometry}
Cardoso~Barbosa,~I.; Gutsche,~J.; Widera,~A. {Impact of charge conversion on NV-center relaxometry}. \emph{Physical Review B} \textbf{2023}, \emph{108}, 075411\relax
\mciteBstWouldAddEndPuncttrue
\mciteSetBstMidEndSepPunct{\mcitedefaultmidpunct}
{\mcitedefaultendpunct}{\mcitedefaultseppunct}\relax
\EndOfBibitem
\bibitem[Alsid \latin{et~al.}(2019)Alsid, Barry, Pham, Schloss, O’Keeffe, Cappellaro, and Braje]{Alsid2019PhotoluminescenceDiamond}
Alsid,~S.~T.; Barry,~J.~F.; Pham,~L.~M.; Schloss,~J.~M.; O’Keeffe,~M.~F.; Cappellaro,~P.; Braje,~D.~A. {Photoluminescence Decomposition Analysis: A Technique to Characterize N-V Creation in Diamond}. \emph{Physical Review Applied} \textbf{2019}, \emph{12}, 044003\relax
\mciteBstWouldAddEndPuncttrue
\mciteSetBstMidEndSepPunct{\mcitedefaultmidpunct}
{\mcitedefaultendpunct}{\mcitedefaultseppunct}\relax
\EndOfBibitem
\end{mcitethebibliography}

\end{document}




\pagebreak


\section{Setup for nanobeam manipulation and scanning measurements}

\begin{figure}[hbt!]
    \centering
    \includegraphics[width=\textwidth]{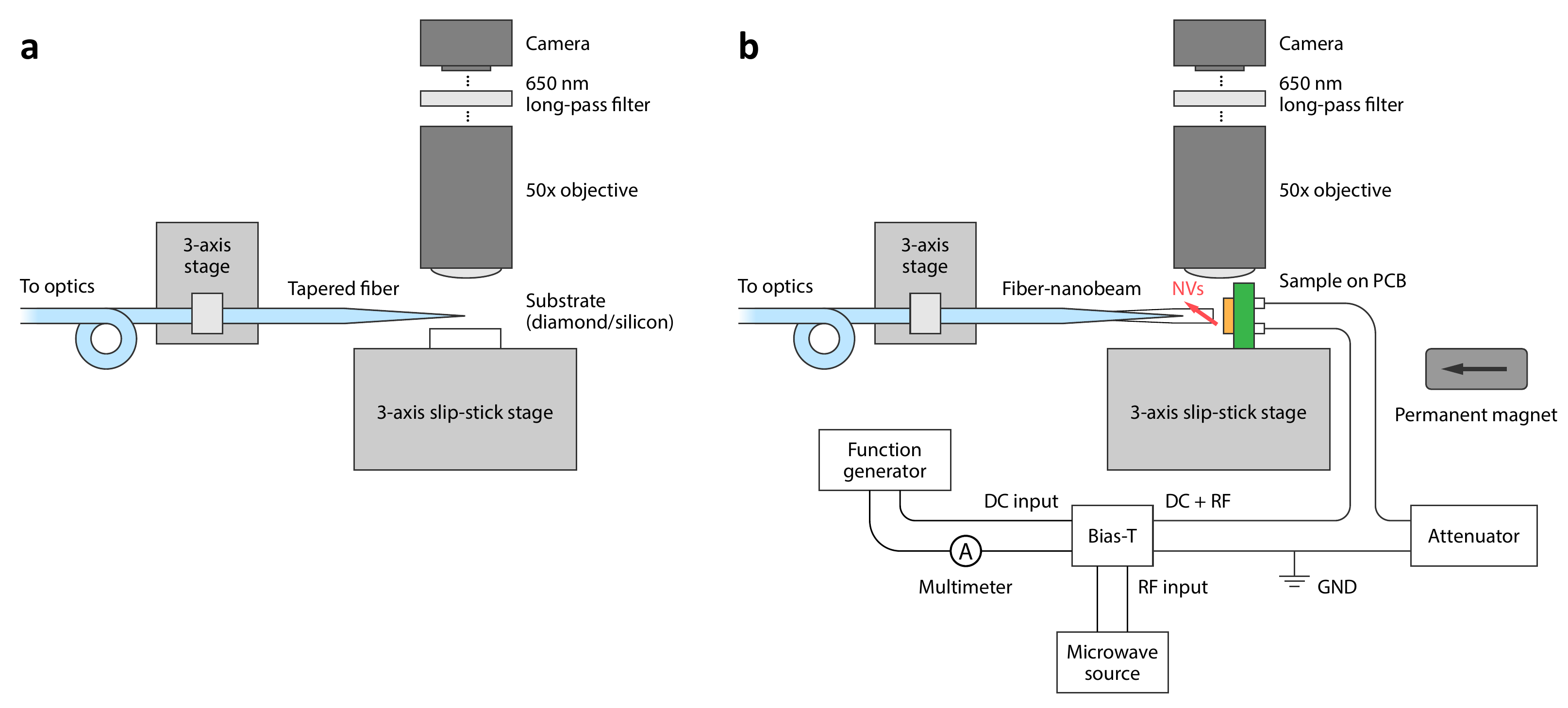}
    \caption{Schematics for the nanobeam manipulation setup, for fiber-nanobeam assembly (a) and scanning measurements (b).}
    \label{fig:setup}
\end{figure}

The setup we use to manipulate the fiber-coupled diamond nanobeams is illustrated in \cref{fig:setup}. A 3-axis slip-stick stage (Mechonics MX-35) is used to position the sample/substrate, and the tapered fiber (and the blunt tip used in the assembly process) is fixed to a separate 3-axis stage for independent coarse positioning. A $50\times$ objective (Mitutoyo M Plan Apo HR, NA = 0.75) provides free-space visual access to the fiber tip and substrate during probe assembly, and can also be used to monitor the probe-sample distance during the scanning measurements when the sample is mounted vertically underneath the objective. A \SI{650}{\nm} long pass filter is included in the imaging system to filter out the excitation laser. 

\section{Analysis on background photoluminescence and the effect of filtering wavelengths}

\begin{figure}[hbt!]
    \centering
    \includegraphics[width=0.6\textwidth]{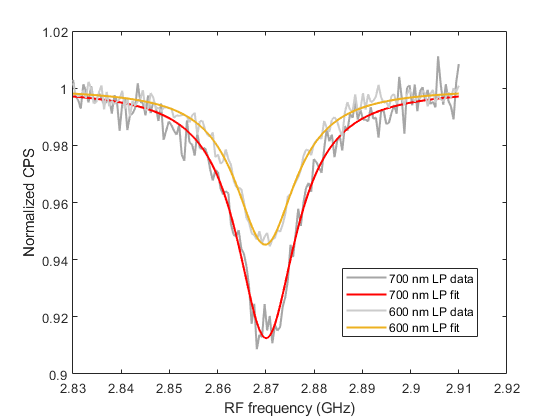}
    \caption{ESR measurement with \SI{600}{\nm} and \SI{700}{\nm} filtering.}
    \label{fig:filters}
\end{figure}

To further identify the composition of the measured photoluminescence, we perform two ESR measurements with the same laser ($P=\SI{11}{\micro\watt}\ll P_\mathrm{sat}$ at fiber input) and microwave power (\SI{19}{dBm} at output), and the same distance between the nanobeam and the microwave strip. The normalized spectra are plotted in \cref{fig:filters}, which sees a 60\% increase in ESR contrast by adding the extra \SI{700}{\nm} filter.

Under the assumption that the background PL from the fiber and the glue are both linear to input laser power, the previous saturation measurement can already exclude their contribution for the low-power measurement. We therefore attribute this increase in contrast to the fluorescence from NV$^0$ centers, which does not contribute to the ESR contrast and has a different spectral distribution. In this case the total PL consists only of the NV$^-$ contribution $I_-$ and NV$^0$ contribution $I_0$, and the ESR contrast can be expressed as
\begin{equation}
    C_{i} = \frac{r C_- I_{-,i}}{r I_{-,i} + (1-r)I_{0,i}}, \quad i=1,2
    \label{eq:contr}
\end{equation}
in which $C_1 = 8.7(3)\%$, $C_2 = 5.5(1)\%$ are the measured contrast under the two filtering regimes, $C_-$ denotes the intrinsic ESR contrast of NV$^-$ independent of filtering, and $r$ is the fraction of NV$^-$ PL in the total PL ($0<r<1$). 

We can calculate the normalized PL intensities $I_-$ and $I_0$ from the spectral density distribution $P(\lambda)$ of NV$^-$ and NV$^0$:
\begin{equation}
    I_{-,i} = \int_{\lambda_i}^\infty P_{-}(\lambda)d\lambda, \quad I_{0,i} = \int_{\lambda_i}^\infty P_{0}(\lambda)d\lambda, \quad i=1,2
\end{equation}
Using the spectrum data from ref. \cite{Savvin2021NVLaser}, we have
\begin{equation}
    I_{-,1} = 0.45(9); \quad I_{-,2} = 0.98(1); \quad I_{0,1} = 0.10(5); \quad I_{0,2} = 0.68(7)
\end{equation}
where the relative uncertainties are estimated from the discrepancy between spectra reported in refs. \cite{Savvin2021NVLaser,CardosoBarbosa2023ImpactRelaxometry,Alsid2019PhotoluminescenceDiamond}, possibly due to the sample and measurement conditions. Substituting these values into \cref{eq:contr} allows us to solve for $C_-$ and $r$:
\begin{equation}
    r = 0.4(2), \quad C_- = 0.12(3).
\end{equation}
The relatively low NV$^-$ concentration obtained here could be because the sidewall NVs are closer to the surface, and are therefore more prone to ionization upon surface defect. The value of $r$ could also be underestimated if part of the non-NV background also has a non-linear power dependence.

\section{Time trace of measured photoluminescence upon nanobeam-sample contact}

\begin{figure}[hbt!]
    \centering
    \includegraphics[width=0.6\textwidth]{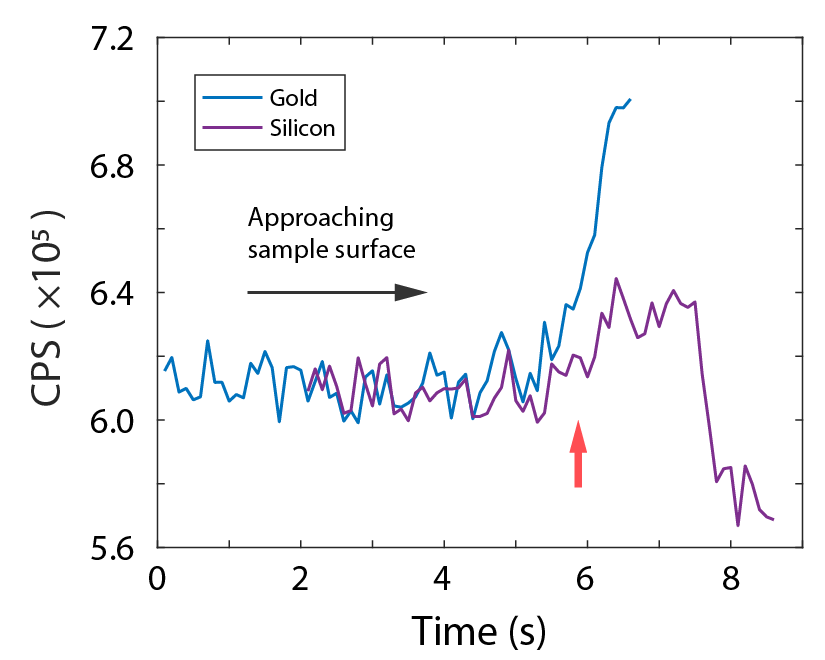}
    \caption{Examples of recorded photoluminescence traces as the fiber-coupled nanobeam approaches a silicon surface and a gold surface. The red arrow indicates the point of contact, marked by an increase in measured counts. For clarity, the two traces have been shifted horizontally so the contact points align.}
    \label{fig:contact}
\end{figure}

When the fiber-coupled diamond nanobeam makes contact with the sample surface, the measured photoluminescence will increase due to reflection from the sample surface. The amplitude of this increase is sensitive to the local reflectivity at the contact point, and can therefore be used to image the sample surface as shown in \cref{fig:contact}. The decrease in the PL trace of silicon after contact results from the decreased coupling efficiency due to the elastic deformation of the fiber-nanobeam assembly. When the deformation is restored upon retracting the probe, the PL counts are also restored. 

\section{Fit results for the 2D scan}

\begin{figure}[hbt!]
    \centering
    \includegraphics[width=0.49\textwidth]{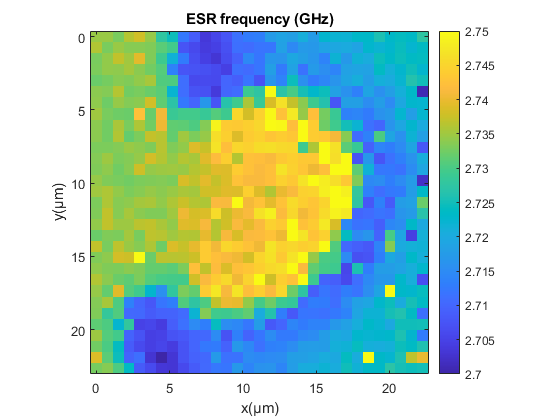}
    \includegraphics[width=0.49\textwidth]{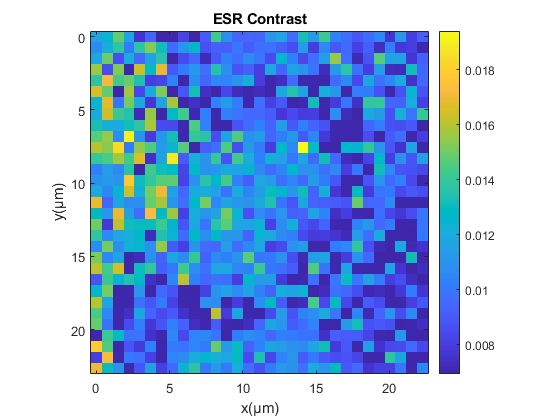}
    \includegraphics[width=0.49\textwidth]{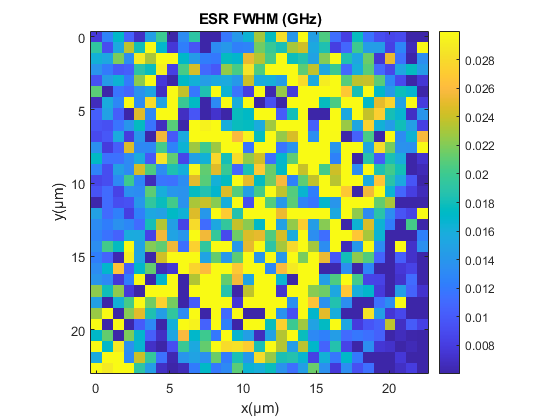}
    \includegraphics[width=0.49\textwidth]{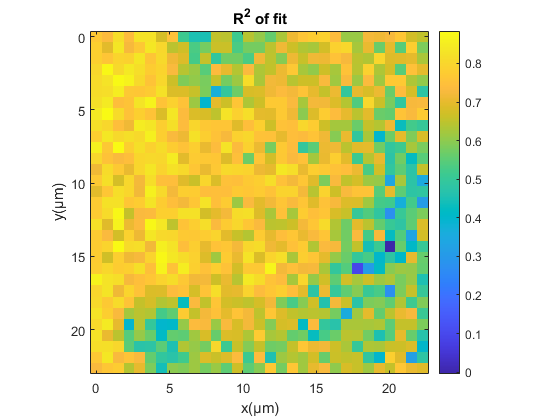}
    \caption{Map of fitting parameters obtained from the 2D imaging data. At each pixel, the frequency of the sidewall NV dip is fixed to $f_0 = \SI{2.729}{\GHz}$, and the fitted frequency (top left), contrast (top right), full width half maximum (bottom left) of the end-facet NV dip is plotted. Bottom right panel plots the $R^2$ of the double Lorentzian fit at each pixel. }
    \label{fig:fitpar}
\end{figure}

\Cref{fig:fitpar} plots the parameters obtained from the double Lorentzian fit of the 2D data in fig.4 of the main text. As discussed in the main text, due to limited signal to noise ratio we take the approximation to fix the frequency of sidewall NV dips to $f_0 = \SI{2.729}{\GHz}$, corresponding to the ESR frequency under the static bias field $B_0 = \SI{5.0}{mT}$ only. The frequency, contrast and width of the ESR dip associated with the end-facet NVs are plotted, along with the $R^2$ parameter of the fit as an indication of the goodness of fit. 

Distortion in the $x$ axis is clearly visible in the plots, as a result of the drift of probe and sample position during the $\sim$ 12-hour scan, executed line by line in $x$. In fig.4 of the main text, this distortion is corrected by manually aligning the left edge of the strip in the contact PL map, and applying the same transformation to the ESR frequency map.

\pagebreak
\bibliography{references}